\documentclass[sn-mathphys-num]{sn-jnl}

\usepackage{graphicx}%
\usepackage{multirow}%
\usepackage{amsmath,amssymb,amsfonts}%
\usepackage{amsthm}%
\usepackage{mathrsfs}%
\usepackage[title]{appendix}%
\usepackage{xcolor}%
\usepackage{textcomp}%
\usepackage{manyfoot}%
\usepackage{booktabs}%
\usepackage{algorithm}%
\usepackage{algorithmicx}%
\usepackage{algpseudocode}%
\usepackage{listings}%

\begin{document}

\title[Article Title]{Born's rule from epistemic assumptions}

\author{\fnm{Per} \sur{\"Ostborn}}

\affil{\orgdiv{Division of Mathematical Physics}, \orgname{Lund University}, \city{Lund}, \postcode{221 00}, \country{Sweden}}

\abstract{
Born's rule is the recipe for calculating probabilities from quantum mechanical amplitudes. There is no generally accepted derivation of Born's rule from first principles. In this paper, it is motivated from assumptions that link the ontological content of a proper physical model to the epistemic conditions of the experimental context. More precisely, it is assumed that all knowable distinctions should correspond to distinctions in a proper model. This principle of \emph{ontological completeness} means, for example, that the probabilistic treatment of the double slit experiment with and without path information should differ. Further, it is assumed that the model should rely only on knowable ontological elements, and that failure to fulfill this principle of \emph{ontological minimalism} gives rise to wrong predictions. Consequently, probabilities should be assigned only to observable experimental outcomes. Also, the method to calculate such probabilities should not rely on the existence of a precise path of the observed object if this path is not knowable. A similar principle was promoted by Max Born, even though he did not apply it to probability. Another crucial assumption is that the proper rule to calculate probabilities should be generally valid. It should be applicable in all experimental contexts, regardless the setup that determines which attributes of the studied object are observed, together with the probability to observe each of the associated attribute values. There is no need to refer to the Hilbert space structure of quantum mechanics in the present treatment. Rather, some elements of this structure emerge from the analysis.}


\maketitle

\section{Introduction}
\label{intro}

Max Born introduced the rule that carries his name in a footnote in the paper \emph{Zur Quantenmechanik der Stoßvorg\"ange} \cite{Born}. According to this rule, if $a$ is an observable value of an attribute $A$, and $\psi(a)$ is an associated complex-valued wave function, it holds that

\begin{equation}
p(a)=|\psi(a)|^{2},
\label{born}
\end{equation}
where $p(a)$ is the probability that a measurement of attribute $A$ yields value $a$.

In his Nobel lecture \cite{bornnobel}, Born stated that Einstein gave him the idea to formulate this rule:

\begin{quote}
\emph{He had tried to make the duality of particles – light quanta or photons – and waves comprehensible by interpreting the square of the optical wave amplitudes as probability density for the occurrence of photons.}
\end{quote}

John von Neumann introduced the Hilbert space formalism of quantum mechanics. When the set $\{a_{j}\}$ of all possible values of $A$ is discrete, it is possible to write

\begin{equation}
|\psi\rangle=\sum_{j}c_{j}|j\rangle,
\label{hilbertstate}
\end{equation}
where $\{|j\rangle\}$ is an orthonormal basis for a Hilbert space associated with attribute $A$, and $\{c_{j}\}$ are complex coordinates in this basis for the state vector $|\psi\rangle$. The basis vector $|j\rangle$ is associated with the attribute value $a_{j}$ via the relation $\bar{A}|j\rangle=a_{j}|j\rangle$, where $\bar{A}$ is the Hermitean operator corresponding to $A$. In this language, Born's rule reads

\begin{equation}
p(a_{j})=|c_{j}|^{2}.
\label{hilbertborn}
\end{equation}

von Neumann also introduced the density operator

\begin{equation}
\bar{D}\equiv\sum_{k}w_{k}|\psi_{k}\rangle\langle\psi_{k}|,
\label{densityop}
\end{equation}
which generalizes the \emph{pure state} described by the state vector $|\psi\rangle$ to a mixture of such states $|\psi_{k}\rangle$, describing a situation where the pure state cannot be known in practice. The statistical weight $w_{k}$ assigned to $|\psi_{k}\rangle$ in the \emph{mixed state} corresponds to the classical probability that the pure state is in fact $|\psi_{k}\rangle$.

In this framework, a generalized Born rule can be formulated as

\begin{equation}
p(a_{j})=\mathrm{Tr}\{\bar{P}_{j}\bar{D}\},
\label{Bornrule}
\end{equation}
where $\bar{P}_{j}$ is the projection operator onto the pure state $|j\rangle$, the subspace of the complex Hilbert space associated with attribute value $a_{j}$, and $\mathrm{Tr}\{X\}$ denotes the trace of $X$.

As far as the author of this paper knows, Born himself did not try to motivate the rule that carries his name theoretically, but referred to experimental `proof' \cite{bornnobel}. Nevertheless, there is a long history of attempts to derive Born's rule from fundamental principles.

In 1932, von Neumann arrived at Born's rule in the form (\ref{Bornrule}) from four assumptions \cite{Neumann}. Two of these assumptions refer to the Hilbert space formalism of quantum mechanics, which is thus assumed. Another assumption states that the expected value $E(A)$ of a measurement of attribute $A$ is linear in the sense that

\begin{equation}
E(\sum_{l}b_{l}A_{l})=\sum_{l}b_{l}E(A_{l}),
\label{neumann}
\end{equation}
where $\{A_{l}\}$ is an arbitary set of attributes, and $\{b_{l}\}$ is an arbitrary set of real constants. Assumption (\ref{neumann}) was criticized by Grete Hermann \cite{Hermann}, and also by John Bell a few decades later \cite{Bell}. The reason is that for sets of attributes whose values are not simultaneously measurable, like position and momentum, it is not clear what $E(\sum_{l}b_{l}A_{l})$ means, or whether this expression can be defined at all. However, it can be defined in quantum mechanics via the relation $\sum_{l}b_{l}\bar{A}_{l}$ between the Hermitean operators $\{\bar{A}_{l}\}$ that correspond to the attributes $\{A_{l}\}$, giving general meaning to the contested assumption (\ref{neumann}). It then holds in quantum mechanics thanks to the specific form of the rule used to calculate expectation values. However, this rule is closely related to Born's rule. Therefore the assumptions used by von Neumann are so closely related to the standard postulates for quantum mechanics that the derivation of Born's rule becomes virtually circular.

Andrew Gleason \cite{Gleason} derived Born's rule in 1957 in the form (\ref{Bornrule}) from the standard Hilbert space formalism of quantum mechanics, together with the assumption that  

\begin{equation}
p(a_{j})=f(\bar{P}_{j}).
\label{noncon}
\end{equation}   
This result is known as Gleason's theorem, and applies in real or complex Hilbert spaces with dimension three or more.

The possibility to express the probability to measure value $a_{j}$ of attribute $A$ according to Eq. (\ref{noncon}) corresponds to an assumption that probabilities in a certain sense are \emph{noncontextual}. Namely, $p(a_{j})$ depends on the projection operator $\bar{P}_{j}$ associated with the quantity value itself, but not on other aspects of the experimental arrangement in which $a_{j}$ is measured.

The critical role of the assumption of noncontextuality in Gleason's theorem is highlighted in a playful paper by Scott Aaronson \cite{Aaronson}, where he explores alternatives to Born's rule. More precisely, he asks whether it is possible to set

\begin{equation}
p_{j}=|c_{j}|^{m}
\end{equation}
for $m>2$ in a pure state $|\psi\rangle$, as expressed in Eq. (\ref{hilbertstate}). It turns out that for $m>2$ a linear evolution $\bar{U}|\psi\rangle$ does not preserve the norm $\sum_{j}|c_{j}|^{m}=1$ except for a very restricted set of evolution operators $\bar{U}$ that correspond to generalized permutation matrices. To make sure all probabilities always add to one, it is, however, possible to set

\begin{equation}
p_{j}=\frac{|c_{j}|^{m}}{\sum_{j}|c_{j}|^{m}}.
\label{conprob}
\end{equation}
These probabilities become \emph{contextual}, since $p_{j}$ depends not only on $\bar{P}_{j}$, which defines $c_{j}$ via $\bar{P}_{j}|\psi\rangle=c_{j}|j\rangle$, but on the entire state $|\psi\rangle$ prior to measurement, which is defined by the experimental arrangement as a whole. Therefore Gleason's theorem does not apply.

The assumption of noncontextuality expressed by Eq. (\ref{noncon}) relates the formalism to the experimental situation. Assumptions of this kind may be called \emph{operational}. They specify conditions under which observations or measurements are made, and translate those into general properties of the physical formalism used to describe these measurements.

From this perspective, Gleason replaced a formal postulate of quantum mechanics (Born's rule) with an operational assumption (noncontextuality). Many researchers have taken a similar route in the last decades, trying to reconstruct quantum mechanics from physical principles relating to the experimental situation or to general rules that govern acquisition of information.

Thomas Galley, Lluis Masanes and Markus M\"uller went one step further than Gleason along that road. They derived Born's rule by replacing not only one but \emph{two} formal postulates of quantum mechanics with operational assumptions \cite{Galley,Masanes}. Keeping the Hilbert space representation of the physical state, these authors removed the measurement postulate, which states that the observation of attribute value $a_{j}$ corresponds to the projection to an updated physical state $\bar{P}_{j}|\psi\rangle=c_{j}|j\rangle$.

As a replacement, they assume that the observation of each value $a_{j}$ of each attribute $A$ that can be measured on the system is associated with a probability $p_{j}=f(a_{j}|\psi)$ such that $\sum_{j}p_{j}=1$ for each state $|\psi\rangle$. Further, they assume that for each system represented by a Hilbert space with finite dimension $D$, there is also a finite set of probabilities

\begin{equation}
\mathbf{p}=\{p_{1},p_{2},\ldots,p_{d}\}
\label{completep}
\end{equation}
for different experimental outcomes such that if this set is known, then the probability $p_{j}$ of any value $q_{j}$ of any observable attribute $A$ can be deduced, given the state $|\psi\rangle$. In effect, the latter assumption means that the state $|\psi\rangle$ of the system is determined by a fixed set $\mathbf{M}$ of measurements and the associated probabilities of the outcomes. The probabilities of the outcome of all other measurements can then be deduced.

This is clearly an operational way to estimate the state, without explicit reference to abstract complex amplitudes $c_{j}$ in a given basis in Hilbert space. Hence the rule how to deduce the probabilities of measurements that are not part of $\mathbf{M}$ does not have to make explicit use of these amplitudes either, but only of the list of probabilities $\mathbf{p}$.

This starting point in the derivation of Born's rule is epistemic in nature, in the sense that it sticks to relations between probabilities of measurements that are possible to perform. The attitude is similar to that of Werner Heisenberg in his paper from 1925 that pioneered the mathematical formulation of quantum mechanics \cite{Heisenberg}.

The power of the operational assumptions used to derive Born's rule is not always spelled out. Sometimes they are regarded as 'natural' or 'self-evident' by the authors, and are therefore not put at the same fundamental level as the standard postulates of quantum mechanics they replace. However, the mere fact that they \emph{can} be used as such replacements shows that they carry crucial weight. For example, the strength of the assumed existence of the set of probabilities $\mathbf{p}$ in Eq. (\ref{completep}) is illustrated by the fact that it excludes Eq. (\ref{conprob}) from the list of possible rules to calculate probabilities, just like Gleason's assumption (\ref{noncon}) of noncontextuality does.

A diminishing attitude to operational assumptions is evident in attempts to derive Born's rule within the framework of the many-worlds interpretation of quantum mechanics (MWI). The obvious reason is that a basic idea of MWI is that the unitary evolution of the state vector in Hilbert space is all there is. Any assumptions used to connect this abstract view to rules that govern the outcome of experiment must therefore be seen as secondary, or be swept under the rug.

In the context of MWI, David Deutsch aimed to derive Born's rule from decision-theoretic assumptions \cite{Deutsch}. Since the evolution in the MWI is deterministic, such assumptions cannot involve probability. Critics have pointed out, however, that even though probability does not enter explicitly in Deutsch' arguments, it is still implicit in the treatment \cite{Barnum}. The starting point of his derivation is an agent trying to bet rationally on the outcome of a game defined by a quantum mechanical superposition.

Charles Sebens and Sean Carroll set ut to derive Born's rule from a slightly different angle \cite{Sebens}. They acknowledged that probability cannot appear from thin air in the MWI. Instead, they argue that probability can be defined after the wave-function split into several new 'branches' when a measurement is made by an apparatus, but before an agent becomes aware of the outcome of this measurement. At this stage, the authors claim that there is a \emph{self-locating uncertainty}, meaning that the agents sitting on each of these new branches cannot know which branch they actually belong to. However, by applying Born's rule they can compute the probability that they will eventually find out that they are located on any given branch. In this way, the probabilities within the MWI become epistemic – they reflect the incomplete knowledge of an aware agent.

Besides the problem how to introduce probability within the MWI, the very use of \emph{agents} as starting points for such attempts is problematic, since there are no fundamental agents in a world consisting of no more than a universal wave function with unitary evolution. One may imagine that it is possible to demonstrate that parts of such a wave function may evolve into a state that can be described as an agent, but in that case a proper derivation of Born's rule from first principles should refer only to the evolving wave function. No extra assumptions should be needed.

In fact, the introduction of agents that make judgments in order to derive Born's rule means that the nature of both the assumptions and the rule that is derived are operational. The Hilbert space formalism recedes into the abstract background. It may be seen as merely a carpet with abstract patterns, on which agents do actual physics, checking probabilities of experimental outcomes.

The problem how to derive the experimental content of quantum mechnics from the abstract framework of the MWI is addressed by Lev Vaidman \cite{Vaidman}. He reviews attempts to derive Born's rule in other approaches to quantum mechanics as well. Vaidman's conclusion is clear: Born's rule cannot be derived from the other postulates of quantum theory without additional assumptions.

The review by Vaidman is more comprehensive than what is offered in this introduction. The focus here is to highlight 
the operational or epistemic content of the needed additional assumptions, and to emphasize that from a logical standpoint they must be seen as fundamental postulates on the same footing as the conventional postulates of quantum mechancis they aim to replace, among those Born's rule.

From this perspective, the essence of many derivations of Born's rule can be seen as attempts to replace formal postulates of quantum mechanics with postulates that can be motivated operationally or epistemically. In this paper, an attempt is made to follow this route to its natural end. Born's rule will be motivated without referring to any standard postulates of quantum mechanics at all. Assumptions that try to encapsulate the epistemic conditions of the experimental situation will suffice.

Figuratively speaking, an attempt is made to pull away the carpet of abstract formalism from under the feet of the agents and their experimental arrangement, and to show that no harm is done. Rather, the Hilbert space formalism tentatively reemerges as a generally applicable algebraic representation of the experimental context itself.

The material presented here is a further development of tentative ideas presented in a wider context in previous papers by the same author \cite{ostborn}. The present paper is organized as follows. Section \ref{defi} introduces concepts and ideas employed in the analysis. The assumptions used to arrive at the desired result are described in section \ref{assu}. They are divided into epistemic assumptions, presented in section \ref{closure}, and conditions that a generally applicable scheme to calculate probabilities must fulfill, listed in section \ref{algebraic}. The motivation of Born's rule is provided in section \ref{motivation}. First, a symbolic formalism that encapsulates the relevant aspects of an experimental context is introduced in section \ref{symbolic}. Then, it is argued in section \ref{algebraicform} that a scheme involving Born's rule is the only acceptable algebraic representation of this formalism, given the assumptions in section \ref{assu}. Section \ref{conclusions} summarizes the main findings, and section \ref{outlook} considers the possibility that the chosen epistemic assumptions may be used to gain clarity on more aspects of physical law than Born's rule.

\section{Ideas and definitions}
\label{defi}

At the core of the present attempt to motivate Born's rule is an epistemic approach to physics. It is assumed that the structure of knowledge and the structure of the physical world reflect each other. A thorough analysis of what can be known about this world and what cannot provides lessons about the world itself. This perspective conditions the choice of definitions and assumptions.

\begin{figure}
\begin{center}
\includegraphics[width=80mm,clip=true]{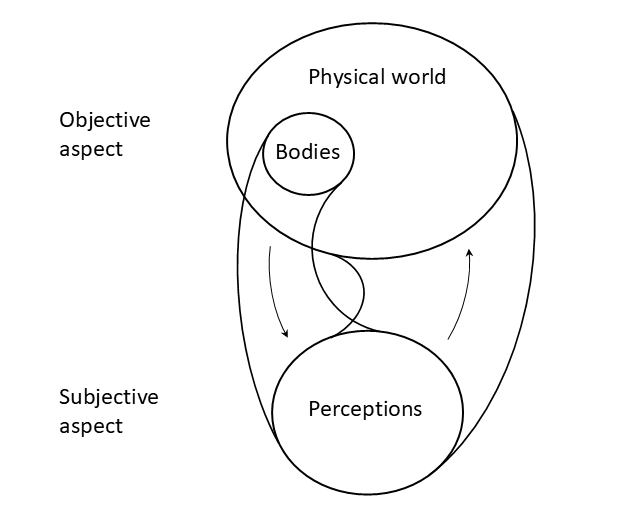}
\end{center}
\caption{Schematic illustration of the world view that inspires the assumptions used to motivate Born's rule. In this view, there is nothing objective in the physical world independent of the subjective, and the subjective is completely embedded in the objective in the sense that all perceptions can be correlated with physical states of bodies of perceiving subjects.}
\label{Figure0}
\end{figure}

Perceiving and knowing subjects are necessary ingredients in such a world view. The relation between the subjective and the objective aspects of the world becomes fundamental, as illustrated in Fig. \ref{Figure0}. It is considered meaningless to talk about perception in itself without any objects to perceive, just as it is empirically meaningless to consider objects that cannot potentially be perceived. The subjective and objective aspects of the world are thus seen as two sides of a coin: distinguishable but inseparable. This model is similar to John Wheeler’s idea of a participatory universe \cite{participatory}.

To be able to speak about and identify objective physical law, like Born's rule, we have to be able to interpret and generalize our subjective perceptions. A fundamental distinction between proper and improper such interpretations has to be assumed. This distinction transcends the perceptions themselves. In this picture, knowledge corresponds to properly interpreted perceptions. To exemplify, a dream is properly interpreted as such, whereas an improper interpretation would be that the perceptions during the dream correspond to objects external to the brain of the dreamer.

\subsection{Potential knowledge}
\label{pknow}

The physical state of the world is taken to correspond to a state of knowledge called the \emph{potential knowledge}, which will be denoted $PK$. The aim of the present discussion about potential knowledge $PK$ is to provide the necessary conceptual background to the use of $PK$ in the following analysis. A more detailed discussion is found in Ref. \cite{ostborn}.

Allowing for potential knowledge to be gained or lost, $PK$ becomes a function of a temporal parameter $n$, that is $PK=PK(n)$. Potential knowledge $PK(n)$ is assumed to be incomplete at all times $n$. The reason is intuitively clear from Figure \ref{Figure0}. This knowledge is represented in the physical state of the bodies of all perceiving subjects, and the set of these bodies is considered to be a proper subset of the physical world. Thus, if $PK(n)$ were complete, this knowledge would be represented in a proper subset of itself. This is not possible under the reasonable assumption that the states of the bodies and the state of the external world can vary independently. Similar hypotheses about the incompleteness of knowledge are reviewed by Szangolies under the poetic headline \emph{epistemic horizons} \cite{Szangolies}. 

The change of $PK$ may be used to \emph{define} the passage of time. If $PK$ does not change at all, there is no epistemic basis for the statement that time has passed. There will be no clock whose pointer is perceived to change position. No thought will pop up in anybody’s head that 'time has passed', since that in itself would amount to a new element of knowledge. These considerations imply that the time $n$ that parameterizes $PK$ is discrete, since it keeps track of a sequence of potentially perceivable changes

\begin{equation}
PK(n)\rightarrow PK(n+1) \rightarrow PK(n+2) \rightarrow \ldots
\end{equation}

To be able to use $PK$ as a basis of a well-defined physical state, it must be possible to define it precisely. However, the knowledge any one of us is consciously aware of at any given time is quite fuzzy. Our attention shifts, we may have vague ideas, our interpretations of what we perceive may be correct or erroneous, or simply not settled. Therefore, a distinction is needed between the \emph{aware knowledge}, which may be fuzzy, and the \emph{potential knowledge}, which is assumed to be precise.

The potential knowledge $PK(n)$ is defined so that it corresponds to all knowledge that may in principle be deduced from all subjective perceptions at time $n$. It includes, for example, experiences at $n$ that we do not become fully aware of or properly interpret until later.

In the present epistemic approach, an \emph{object} may simply be defined as an element of perception, which can be subjectively distinguished from other such elements. Knowledge about objects may be defined as proper interpretation of elements of perceptions. The distinction between knowledge and sense perceptions makes it possible to speak of objects that are hidden at times, such as the sun at night, or microscopic and abstract objects like elementary particles, which are deduced from a systematic analysis of macroscopic observations.

The entity $PK(n)$ is assumed to consist of potential knowledge about a set of objects and their attributes. An \emph{attribute} $A$ can be defined as a set of qualities $\{a\}$ of an object that can be associated with each other and ordered. One quality in such a set can be called an attribute value $a$. Red, green and blue are three different such qualities or attribute values that can be ordered into a spectrum that defines the attribute \emph{color}. Attributes may be internal or relational. An internal attribute, such as color, refers to the object itself, whereas a relational attribute, such as distance or angle, relates two or more objects.

The existence of several subjects is assumed, and the potential knowledge $PK(n)$ is assumed to be a collective entity that amounts to the union of the potential knowledge at time $n$ of all perceiving subjects. The possibility is assumed that different subjects observe the same object, that is, that they share elements of perception. This circumstance may be considered as a definition of the statement that these subjects are part of the same world. The collective nature of $PK(n)$ makes it possible to define a corresponding epistemic physical state that transcends the perspective and perceiving power of individual observers, and thus may become a proper representation of the physical world as a whole.

The underlying assumption that different subjects may share perceptions is seen as fundamental, meaning that it cannot be fully accounted for in terms of something else. The most common attempted such explanation is that hypothesized observer-independent objects emit electromagnetic radiation that enters the eyes of different observers triggering a common perception. Even though such observer-independent objects are at odds with the world view proposed here, the present model provides objectivity to perceptions in the sense that they transcend the individual perceiving subject when they are shared, just as the proper interpretation of such shared perceptions is assumed to transcend the individual interpreting subject, allowing for common knowledge.      

It should be stressed that potential knowledge at the conceptual level is considered to be independent of any particular representation or encoding of this knowledge, such as formal propositions or strings of bits. The possibility to distinguish knowledge from its representations follows from the assumption that subjective perceptions and insights about these perceptions are fundamental notions.

The epistemic physical state $S(n)$ introduced in Ref. \cite{ostborn} is an example of a representation of $PK(n)$. This state is not needed in the present study, but it can be seen as a generalisation of the set $S_{A}$ to be discussed in the next section.    

\subsection{Alternatives}
\label{alternatives}

Born's rule will be associated with probabilities of a set of possible updates of potential knowledge $PK(n)\rightarrow PK(n+1)$, where the update corresponds to a gain of potential knowledge of the value $a$ of an attribute $A$ of some object. Such a gain of knowledge may be expressed as a reduction of the set of attribute values $S_{A}(n)=\{a\}$ that are not excluded by potential knowledge at time $n$ according to $S_{A}(n+1)=SR_{A}$, where $SR_{A}\subset S_{A}(n)$.

Given that $PK(n)$ is argued to be incomplete, several different such updates $S_{A}(n)\rightarrow SR_{A}$ or $S_{A}(n)\rightarrow SR_{A}'$ may be possible, meaning that there is no potential knowledge at time $n$ that makes it possible to predict which of them will occur, if any. This is an expression of the fundamental indeterminism that follows from the association of the physical state with a state of incomplete knowledge. Each of the possible reduced sets of attribute values $SR_{A}$ or $SR_{A}'$ may be called an \emph{alternative}.

A \emph{complete set of alternatives} $CS_{A}$ at time $n$ with respect to attribute $A$ of a given object may be defined as a set $CS_{A}=\{SR_{A}\}$ such that $\bigcup SR_{A}=S_{A}(n)$, $SR_{A}\subset S_{A}(n)$ for each element in the set, and $SR_{A}\cap SR_{A}'=\varnothing$ for any two such elements. Simply put, a complete set of alternatives is an exhaustive collection of mutually exclusive ways in which potential knowledge of an attribute $A$ may be expanded.

An individual alternative in the set $CS_{A}$ may be labelled $A_{j}$. According to the above discussion, it is a set that has one or several attribute values $a$ as elements. Since any two alternatives in $CS_{A}=\{A_{j}\}$ are subjectively distinguishable by definition, they must form a countable or discrete set.

The prospect that one of the alternatives $A_{j}$ in a complete set $\{A_{j}\}$ will be revealed at some later time $n+m$ may be classified according to Table \ref{levels}. Which of these \emph{knowability levels} apply is a function of $PK(n)$. This is so since $PK(n)$ is assumed to correspond to the physical state at time $n$. All that can be said at that time about the future physical state therefore follows from applying physical law to the evolution of $PK(n)$, including the knowability level of a given set $\{A_{j}\}$. 

\begin{table}
\caption{Three knowability levels of complete sets of alternatives $\{A_{j}\}$}
\label{levels}
\begin{tabular}{ll}
\hline\noalign{\smallskip}
		1 & \emph{No alternative will ever be known to come true}. There is no $m>0$ and no\\
		  & element $A_{j}$ in the set such that it may hold that $S_{A}(n+m)=A_{j}$.\\
\noalign{\smallskip}
		2 & \emph{It may become known which alternative is true}. There is a $m>0$ such that it\\
		  & may hold that $S_{A}(n+m)=A_{j}$ for some element in the set, but it is not\\
		  & dictated by physical law that this will become true.\\
\noalign{\smallskip}
		3 & \emph{It will become known which alternative is true}. There is also a $m'\geq m$ such\\
		  & that physical law dictates that $S_{A}(n+m')=A_{j}$ for some element in the set.\\
\noalign{\smallskip}\hline
\end{tabular}
\end{table}

\subsection{Experimental contexts}
\label{contexts}

Probabilities emerging from Born's rule will be treated as functions of the entire \emph{experimental context}, rather than of the object studied in this context, even though the probabilities refer to this object. This is the only consistent approach given the epistemic ansatz. This approach makes it meaningless to speak about probabilities that do not correspond to observations, and all observations takes place in some macroscopic context. To speak about the probability that something will happen to an object without any predefined means to observe this outcome is considered meaningless. An example would be the timing of a radioactive decay of an atomic nucleus situated in a distant star.

An experimental context $C$ consists of several parts. The observed object $O$ and the body of an observer $OB$ are necessary components in $C$. Object $O$ may be divided into a \emph{specimen} $OS$ and an \emph{apparatus} $OA$, with which values of attributes of the specimen of interest are measured. When the specimen $OS$ is deduced rather than observed directly, like an electron, the apparatus can be divided into a \emph{machine} $OM$ and a \emph{detector} $OD$, where observed changes in the state of the detector define the outcome of the experiment. For example, in a double-slit experiment, the specimen $OS$ is a particle ejected from a gun. This gun, together with the slits and supportive structure constitute the machine $OM$. The detector $OD$ is the screen on which the particle ends up.

In the state of potential knowledge $PK_{C}(n_{i})$ of an experimental context $C$ at the time $n_{i}$ the experiment starts, there should be knowledge about which states of the detector $OD$ corresponds to which states of the specimen $OS$. In conventional language, the states of the specimen and detector should be entangled. There should also be knowledge about one or several complete sets of alternatives $\{A_{j}\}$ that apply to attributes of the specimen $OS$. At least one of these sets should have knowability level 3 at time $n_{i}$ according to Table \ref{levels}, meaning that a proper context $C$ should always produce a definite outcome once the experiment is started. The end of the experiment at time $n_{f}$ is defined as the observation of an attribute to which such a complete set of alternatives at knowability level 3 is associated at time $n_{i}$.

The body $OB$ of an observer has finite size and finite resolution power, just like an apparatus $OA$  Therefore, the number of alternatives $A_{j}$ in each complete set $\{A_{j}\}$ defined within $C$ must be finite.

As an example, in the double-slit experiment there is a complete set of alternatives $\{A_{j}\}=\{A_{1},A_{2}\}$ that corresponds to the passage of the specimen through slit 1 or slit 2. This set of alternatives has knowability level 1 if there is no detector at the slits. Another set of alternatives $\{A_{j'}'\}$ is defined by the detector screen, where the number of alternative positions of the specimen depends on the resolution of the detector.

\begin{figure}[tp]
\begin{center}
\includegraphics[width=100mm,clip=true]{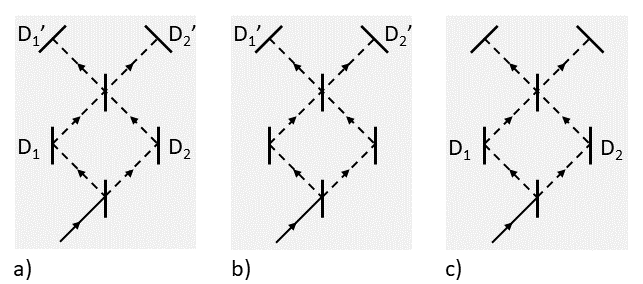}
\end{center}
\caption{Three experimental contexts $C$ corresponding to an adjustable Mach-Zehnder interferometer. Attribute $A$ with two possible alternatives $A_{1}$ or $A_{2}$ corresponds to the passage of the left or right mirror, whereas attribute $A'$ with alternatives $A_{1}'$ or $A_{2}'$ corresponds to the final absorption to the left or right. Both these attributes correspond to subsequent positions of the specimen $OS$. The presence of a detector to decide the value of the attribute is marked by the letter $\mathrm{D}$. a) Two complete sets of alternatives $\{A_{j}\}$ and $\{A_{j'}'\}$ are defined, both at knowability level 3. b) The set of alternatives $\{A_{j}\}$ is degraded to knowability 1, meaning that it will be forever impossible to know which mirror the specimen passed. c) The set of alternatives $\{A_{j'}'\}$ is degraded to knowability 1, meaning that it will be forever impossible to know where the specimen ends up. }
\label{Figure1}
\end{figure}

Another example is given by the Mach-Zehnder interferometer shown in Fig. \ref{Figure1}. In that case $\{A_{j}\}$ corresponds to the two alternatives $\{A_{1},A_{2}\}$ that the specimen hits the left or right mirror, and $\{A_{j'}'\}$ corresponds two the two alternatives $\{A_{1}',A_{2}'\}$ that it ends up in the left detector $D_{1}'$ or the right detector $D_{2}'$.

It is allowed that the knowability level of some complete set of alternatives may change during the course of the experiment. This is the idea behind \emph{delayed-choice experiments}, introduced by John Wheeler \cite{delayed}. If the context $C$ is such that the second beam splitter along the path of the specimen in the context shown in panel b) of Fig. \ref{Figure1} may be removed after the specimen has passed the mirrors, then knowledge will be gained which mirror it was reflected from when it hits one of the detectors, since detection to the left means that it passed the right mirror, and vice versa. In effect, the removal of the beam splitter means that the knowability level of $\{A_{j}\}$ rises from 2 to 3. It is also conceivable that the knowability level declines from 2 to 1 after the start of the experiment. Such a senario may be realized by means of so called \emph{quantum erasers}, which irreversibly removes knowledge about the path of the specimen \cite{eraser1,eraser2,eraser3}.

\subsection{Probabilities and propensities}
\label{probprop}

As outlined above, probability $p_{j}$ is considered to be an operational quantity that is defined within an experimental context $C$, where it may apply to alternatives $A_{j}$ in a complete set $\{A_{j}\}$ at knowability level 3 according to Table \ref{levels}. According to the discussion in the preceding section, when such probabilities are defined it holds that $p_{j}=f_{j}[PK_{C}(n_{i})]$, but $p_{j}\neq f_{j}[PK_{OS}(n_{i})]$, where $OS$ is the specimen having the attribute $A$. In this way, probabilities become attributes of the context $C$ as a whole, seen as a single object.

There may be contexts $C$ with well-defined complete sets of alternatives $\{A_{j}\}$ for which corresponding probabilities $p_{j}$ are not part of $PK_{C}(n_{i})$ since the details of the experimental setup are not sufficiently known to be able to deduce them from physical law. Such probabilities may be estimated in the frequentist sense if the experiment is such that it can be repeated many times. Note that after such a sequence of experiments we have a new context $C'$ with additional potential knowledge attached, meaning that $PK_{C}(n_{i})\subset PK_{C'}(n_{i}')$, where $n_{i}'>n_{i}$.

When probabilities are potentially known without the need to perform repeated experiments, they correspond to \emph{propensities}. An example would be the probability of each outcome when a dice is thrown, which are known beforehand because of the symmetry of the dice. Another example is the probability that ionizing radiation from a radioactive sample of known type will be detected within a given time frame. This probability can be deduced from physical law.

From the present philosophical perspective, propensities $\pi_{j}$ are preferably defined within an experimental context $C$, just like probabilities, where they may apply to alternatives in a complete set $\{A_{j}\}$. However, in contrast to probabilities, they can be defined regardless the knowability level of $\{A_{j}\}$. It is not necessary to actually throw a dice to deduce that the propensity for each outcome is $1/6$. This means that just like a probability $p_{j}$ may be a propensity $\pi_{j}$, but this is not always true, a propensity may be a probability, but that is not always true either. The situation for which it will be argued that Born's rule is needed is when there are probabilities that are conditioned on propensities, which are not probabilities themselves.

Karl Popper put forward the \emph{propensity interpretation of probability}, in which probabilities are seen as properties inherent in the physical system that generates the outcome to which these probabilities are assigned \cite{popper}. He came to this view thinking about the quantum mechanical double-slit experiment, which convinced him that probabilities must be 'physically real'. To clarify what he meant by this, Popper wrote:

\begin{quote}
\emph{They are not properties inherent in the die, or in the penny, but in something a little more abstract, even though physically real: they are relational properties of the experimental arrangement – of the conditions we intend to keep constant during repetition.}
\end{quote}

This view is similar to that assumed in this paper, in which probabilities and propensities are seen as functions of the state of a single experimental context $C$.

However, Popper rejected the epistemic approach that identifies this state with a state of knowledge. In fact, he seems to have argued that such an approach, which underlies the Copenhagen interpretation of quantum mechanics, implies that probabilities become subjective, meaning that they reflect the personal degree of belief of the experimenter, as quantified by her willingness to bet on specific outcomes at given odds \cite{ramsey, definetti}. Indeed, this view is a defining property of the so called \emph{quantum Bayesian interpretation of quantum mechanics}, often called \emph{QBism}, which can be seen as a modern offspring of the Copenhagen interpretation \cite{mermin}.

The present approach to probability and propensity is different, combining subjective and and objective elements. Probability is seen as subjective in the sense that it is predicated on an experimental context $C$, which always contains a subjective observer and is described by an epistemic state. However, it is objective in the sense that once the context $C$ is given, the observer has no say on the probabilities of the observations that are possible within this context. These are functions of the state of potential knowledge $PK_{C}(n_{i})$, which transcends the judgment of a single observer in several ways. It relies on the distinction between correct and erroneous interpretations of perceptions, it takes into account not only the knowledge an observer is consciously aware of, but all knowledge she may acquire in principle from perceptions at time $n_{i}$, and it sums up the potential knowledge of all observers.

In a similar manner, Wolfgang Pauli emphasized the objective aspect of probabilistic measurements in quantum mechanics, even though he, broadly speaking, subscribed to the Copenhagen interpretation \cite{pauli}:

\begin{quote}
\emph{[T]here remains still in the new kind of theory an objective reality, inasmuch as these theories deny any possibility for the observer to influence the results of a measurement, once the experimental arrangement is chosen. Therefore particular qualities of an individual observer do not enter the conceptual framework of the theory.}
\end{quote} 

\section{Assumptions}
\label{assu}

The philosophical ideas introduced in section \ref{defi} are presented in as much detail as necessary to make it understandable why the author of this paper has chosen the following assumptions, but not in any more detail than that. These assumptions provide the foundation of the present study, not the underlying philosophy that has inspired them.

\subsection{Ontological closure}
\label{closure}
The idea that the structure of knowledge and the structure of the world reflect each other means that there should be a one-to-one correspondence between forms of appearances and degrees of freedom in proper physical models. Here \emph{forms of appearances} are understood in the Kantian sense as categories of perception that are given \emph{a priori}, into which all sensory impressions and all other elements of consciousness are placed. The idea that the forms of appearances and the structure of proper physical models reflect each other conforms to the fact that degrees of freedom in physical models and their evolution laws are given \emph{a priori}, just like forms of appearance.

\begin{figure}[tp]
\begin{center}
\includegraphics[width=80mm,clip=true]{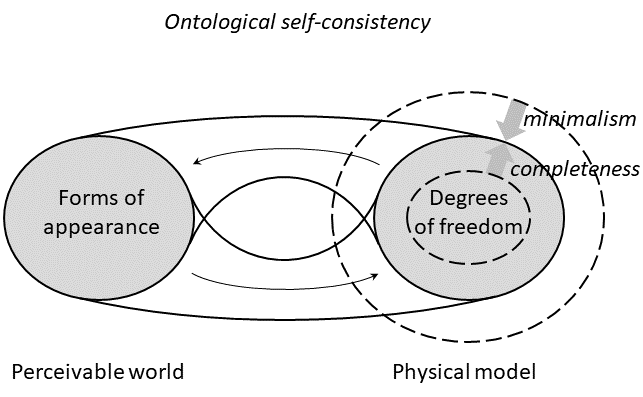}
\end{center}
\caption{It is assumed that a proper physical model should be complete, meaning that it contains all knowable degrees of freedom. The model should also be minimal, meaning that it does not contain any other degrees of freedom. All degrees of freedom of the model should follow from the forms of appearances, and it should be possible to derive all forms of appearances from the model. Such a model may be called ontologically self-consistent. It is argued that a model that is not ontologically self-consistent is insufficient or wrong. Such a loop of self-consisteny reflects the loop that describes the relation between the subjective and objective aspects of the world according to Fig. \ref{Figure0}.}
\label{Figure2}
\end{figure}

This idea can be separated into two principles that limit the set of conceivable physical models. On the one hand, the model should be complete in the sense that it explicitly contains symbolic counterparts of all fundamental forms of appearance. In other words, all knowable degrees of freedom should correspond to independent mathematical degrees of freedom in the model. On the other hand, a proper model should be minimal in the sense that it does not contain any additional such degrees of freedom. If both of these conditions are fulfilled, the model may be called ontologically self-consistent, as illustrated in Fig. \ref{Figure2}.

These principles are somewhat vague. They try to express something very general in few words, thereby resorting to terms them are not precisely defined. To be able to use the principles in order to choose proper models, they will be expressed in a simpler, but also more limited form below, which express the skeleton of the idea. If both of the below conditions below are fulfilled, the model is said to possess \emph{ontological closure}.

\subsubsection{Ontological completeness}
The condition of \emph{ontological completeness} amounts to an assumption that all subjectively knowable distinctions correspond to distinctions in proper physical models. Failure to fulfill this principle in a proposed physical model is assumed to lead to inability to account for all experimental facts, or to wrong predictions.

As an example, attempts to model the physical world as a collection of bits or qubits according to the catchphrase \emph{It from bit} are assumed to fail, since such models do not distinguish between attributes of a given object that have binary values from attributes that can attain more values, or those that have a continuous set of values. Such distinctions are evident from the subjective point of view.

According to the chosen epistemic ansatz, the knowable world at any time $n$ can be divided into two parts by a moving curtain of ignorance: the part of the world about which we currently have potential knowledge $PK(n)$, and the part of the world about which we have no such knowledge. This distinction should be honored in physical models, meaning that the behaviour of objects or attributes that are part of $PK(n)$ should be modelled differently than those that are not. Specifically, it means that a double-slit experiment $C$ where information about which slit the specimen passes is forever outside potential knowledge should be described and behave differently than a corresponding experimental context $C'$ where such path information will become known. In either case, the knowable dynamics corresponds to procedures to calculate probabilities $p_{j}$ for different outcomes $A_{j}$. These procedures should therefore differ in contexts $C$ and $C'$.

The same distinction should be made between the two Mach-Zehnder interferometer contexts in Fig. \ref{Figure1}(a) and \ref{Figure1}(b). In context (a), with path information, the classical rules of probability apply, giving

\begin{equation}
p_{j'}'=p_{1}p_{1j'}+p_{2}p_{2j'},
\label{classicalp}
\end{equation}
where $p_{j}$ is the probability that the specimen passes mirror $j$, $p_{jj'}$ is the probability that it arrives at detector $D_{j'}$ given that it passes mirror $j$, and $p_{j'}'$ is the probability that it hits detector $D_{j'}'$. In context (b), without path information, according to the principle of ontological completeness, it must be required that

\begin{equation}
p_{j'}'\neq\pi_{1}\pi_{1j'}+\pi_{2}\pi_{2j'}.
\label{quantump}
\end{equation}  
Note that the quantities at the right hand side of this equations are propensities that no longer are probabilities, according to the discussion in section \ref{probprop}. In a very general sense, Eq. (\ref{quantump}) expresses the necessity of \emph{interference}, since $p_{j'}'$ cannot be calculated \emph{as if} there were two independent, well-defined paths, one of which is chosen by the specimen.

\subsubsection{Ontological minimalism}
\label{minimalism}
The condition of \emph{ontological minimalism} corresponds to the assumption that all distinctions in proper physical models correspond to subjectively knowable distinctions. Failure to exclude in a model distinctions that are not knowable is assumed to lead to wrong predictions.

It is a general lesson from the development of physics in the twentieth century that ontological dead weight in physical models should be thrown overboard. There turned out to be no knowable consequences of the introduction of the aether; therefore it was abandoned. More than that, an aether that defines an absolute spatial rest frame is inconsistent with the empirically confirmed predictions of special relativity. In a similar manner, wrong predictions are obtained when operations with no knowable effect are taken into account in physical models, such as the interchange of two identical particles in statistical mechanics, or the rotation of a spherically symmetric object. Physical law is ontologically picky.

In the same vein, the condition implies that it will give rise to wrong predictions to make use of path information in a model of a double-slit experiment or a Mach-Zehnder interferometer when in reality there is none ever to be gained. This is so since the distinction between the statements "it follows this path" or "it follows that path" is unknowable. Therefore, the specimen should not be treated \emph{as if} it follows either path. Again, Eq. (\ref{classicalp}) is excluded, since it relies on conditional probabilities for a given path.

The gist of ontological minimalism was identified long time ago. It guided Heisenberg in his original formulation of quantum mechanics \cite{Heisenberg}. Born sharpened and generalized the idea. In his Nobel lecture \cite{bornnobel}, he said:

\begin{quote}
\emph{The principle states that concepts and representations that do not correspond to physically observable facts are not to be used in theoretical description.}
\end{quote}

In a paper dedicated to Niels Bohr on the occasion of his 70th birthday \cite{bornbohr}, Born expands on the subject:

\begin{quote}
\emph{This principle was certainly operative in many instances since Newton's time. The most glaringly successful cases are Einstein's foundation of special relativity based on the rejection of the concept of aether as a substance absolutely at rest, and Heisenberg's foundation of quantum mechanics based on the elimination of orbital radii and frequencies of electronic structures in atoms. I think that this principle should be applied also to the idea of physical continuity.}
\end{quote}

If there is anything new to the ontological minimalism used in this paper, except for the name, it is the idea that any deviation from this principle in physical models leads to wrong predictions. This is a stronger statement than just to say that such ontological dead weight is unnecessary. The present, stronger version of the principle means that it can be used to discriminate between candidate physical models, for instance between conceivable rules to calculate probability. As far as the author of the present paper knows, Born did not try to use this principle to motivate the rule to calculate probabilites that carries his name.

However, as the quote makes clear, Born tried to apply the principle to the treatment of continuity. He argued that it lacks physical meaning to state that the position $x$ of some particle is, say, $x=\pi$ or $x=\sqrt{2}$. To verify this statement would require a measurement with infinite precision. This is impossible in principle, since all measurement devices have finite resolution power. Therefore, Born argued, such statements should be eliminated from physics. Instead, he suggested that a probability density $p(x)$ centered around $x=\pi$ or $x=\sqrt{2}$ should be used.

It may be objected, though, that the use in physics of a density $p(a): \mathbb{R}\rightarrow\mathbb{R}$ in itself violates ontological minimalism. It presupposes a precise probability $p(a)da$ that the attribute value $a$ value will be found in the interval $[a,a+da)$, where $da$ may be arbitrary small, despite the fact that there is no possible experience corresponding to this statement.

The approach in this paper seems more consistent in this respect. As described in section \ref{probprop}, probabilities are only assigned to alternatives $A_{j}$ in a complete set $\{A_{j}\}$ with a finite set of elements, for which it is known beforehand that one of the alternatives $A_{j}$ will come true. Each alternative $A_{j}$ may correspond to a continuity of possible values $A_{j}=\{a\}$ that cannot be excluded given an incomplete potential knowledge $PK_{C}$ about the experimental context $C$. However, the precise values $a$ never appear explicitly in the formalism that is developed. Referring to ontological minimalism, it is also prescribed that the rule to calculate these probabilities from the state of the experimental context should only be applied to such alternatives at a point in time when they have a chance to be realized.

\subsection{Algebraic expression of physical law}
\label{algebraic}

Apart from the principle of ontological closure, a second fundamental assumption is that there is a generally applicable algebraic expression of physical law. In the present setting, this means that it should be possible to find an algebraic scheme with an associated rule to calculate probabilities that is applicable in a wide class of experimental contexts $C$.

In this section, this class of contexts is delineated. Also, a list of conditions is presented, which must be fulfilled by such an algebraic scheme in order to be generally applicable within this class.

\subsubsection{The relevant class of experimental contexts}
\label{contextclass}
Interest is restricted to contexts $C$ for which propensities for all alternatives in all complete sets of alternatives defined within $C$ are part of potential knowledge $PK_{C}(n_{i})$. Conditional propensities $\pi_{jj'}$ to see alternative $A_{j}'$ given a preceding alternative $A_{j}$ is also assumed to be part of $PK_{C}(n_{i})$. If the complete sets of alternatives $\{A_{j}\}$ and $\{A_{j'}'\}$ both have knowability level 3 according to Table \ref{levels}, these conditional propensities are also probabilities $p_{jj'}$. A selection of contexts of this type is shown in Fig. \ref{Figure3}, represented as directed networks of alternatives.

The restriction that all propensities are potentially known at the start of the experiment means that it is possible to deduce them from physical law from $PK_{C}(n_{i})$. This does not imply that they are part of the aware knowledge of the experimenters. 

\subsubsection{The scheme should apply at all times}
\label{alltimes}
The same rule to calculate probabilities should be valid at all times $n$ in the interval $n_{i}\leq n < n_{f}$, where $n_{i}$ is the time at which the experiment starts, and $n_{f}$ is the time at which it ends, meaning that the update $PK_{C}(n_{f}-1)\rightarrow PK_{C}(n_{f})$ is defined by the observation of the attribute value that defines the outcome of the final complete set of alternatives defined within the experimental context $C$.

This requirement can be seen as an expression of the freedom, in principle, to change the timing of observations of different attributes during the experiment. 

\subsubsection{The scheme should apply for all kinds of measurements}
\label{allalt}
The same rule to calculate probabilities should be valid regardless the number of attributes $A$ that are observed within $C$, with corresponding complete sets of alternatives $\{A_{j}\}$, and regardless the number $M$ of alternatives $A_{j}$ within each of these sets.

Figure \ref{Figure3} shows four kinds of contexts where the number of attributes that are observed vary from two to three, and so does the the number of alternative outcomes in each of these observations. A proper algebraic scheme should cover all these cases, with the same rule to calculate probabilities in each. 

\subsubsection{The scheme should apply regardless knowability levels of alternatives}
\label{allknow}
The same rule to calculate probabilities should be valid regardless the knowability levels at the start of the experiment, at time $n_{i}$, of the complete sets of alternatives defined within $C$, and regardless whether any alternatives that are part of such sets having knowability level 2 at time $n_{i}$ actually come true during the course of the experiment, at some time $n$ in the interval $n_{i}<n<n_{f}$.

\subsubsection{The scheme should apply for all sets of propensities}
\label{allprob}
For contexts $C$ with a given set $\{\{A_{j}\},\{A_{j'}'\},\ldots\}$ of complete sets of alternatives, with given associated knowability levels, there is experimental freedom to adjust the parameters that describe $C$ so that propensities for each alternative in each complete set can be varied independently.

For the first attribute $A$ defined within $C$, for which there are $M$ alternative outcomes $\{A_{1},A_{2},\ldots,A_{M}\}$ in the complete set $\{A_{j}\}$, there are $M-1$ independent propensities $\pi_{j}$. For the next attribute $A'$ defined within $C$, there are $M(M'-1)$ independent conditional propensities $\pi_{jj'}$, quantifying the propensity for alternative $A_{j'}$ given alternative $A_{j}$. These $M(M'-1)$ propensities are real numbers in the interval $[0,1]$.

The rule to calculate probabilities must respect this parametric freedom, so that it can be applied to all choices of $\{\pi_{j}\}$ and $\{\pi_{jj'}\}$. If there is another attribute $A''$ defined within $C$ with an associated complete set of alternatives $\{A_{j''}''\}$, the same goes for the $M'(M''-1)$ propensities $\{\pi_{j'j''}\}$, and so on.

\subsubsection{The scheme should respect the independence of sequential propensities}
\label{independent}
The experimental freedom expressed in the preceding section means that it should be possible to choose parameters in the algebraic scheme that describe propensities for the alternatives in a complete set $\{A_{j'}'\}$ independently from the parameters that describe a preceding complete set of alternatives $\{A_{j}\}$.  

\begin{figure}[tp]
\begin{center}
\includegraphics[width=80mm,clip=true]{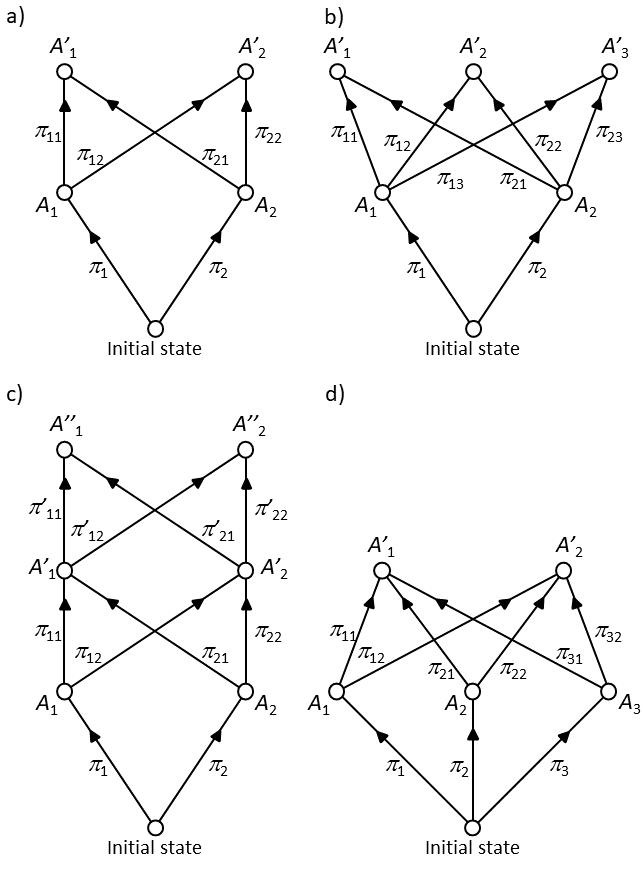}
\end{center}
\caption{Experimental contexts $C$ depicted as networks of alternatives. The nodes in each row correspond to alternatives $A_{j}$ associated with a given attribute $A$. Time flows upwards, as indicated by the directed edges. Propensities $\pi$ can be seen as relations between alternatives. The propensities for the last set of alternatives are always probabilities $p$, since an experiment always ends with a measurement. Different knowability levels according to Table \ref{levels} are not distinguished here.} 
\label{Figure3}
\end{figure}

\section{Motivation of Born's rule}
\label{motivation}

\subsection{Symbolic formalism}
\label{symbolic}
An algebraic scheme that is able to describe all experimental contexts $C$ in the class described in section \ref{contextclass} should contain quantities that represent individual probabilities or propensities $p_{j}$, $p_{jj'}$, $\pi_{j}$, or $\pi_{jj'}$. Since the same scheme should apply at all times according to section \ref{alltimes}, and regardless the knowability levels of the involved sets of alternatives according to section \ref{allknow}, the same quantities $c_{j}$ and $c_{jj'}$ should be used in all cases to represent propensities, regardless whether they are also probabilities.

To each alternative $A_{j}$ in each complete set $\{A_{j}\}$ defined within $C$, it is assumed that there is an associated quantity $c_{j}$ in the algebraic scheme that represents the propensity for $A_{j}$. The association will be expressed as

\begin{equation}
\left(c_{j}|A_{j}\right).
\label{association}
\end{equation}
Interest is restricted further to algebraic schemes such that there is a function $f$ which gives the probability for $A_{j}$ as

\begin{equation}
p_{j}=f(c_{j}).
\label{pf}
\end{equation}
The function $f$ thus connects the algebraic scheme to the empirical data. It should be stressed that the quantity $f(c_{j})$ represents a probability if and only if $\{A_{j}\}$ has knowability level 3.  

To proceed, it may be helpful to introduce a time-dependent symbolic representation of the epistemic state of a context $C$, using associations like that in Eq. (\ref{association}) as elements. Such a representation may be denoted $\overline{PK_{C}}(n)$. After that, the main task will be to choose generally applicable algebraic relations between the involved symbols, and to find a generally applicable function $f$.

For simplicity, consider first the Mach-Zehnder interferometer illustrated in Fig. \ref{Figure1}, represented as a network of alternatives in Fig. \ref{Figure3}(a). The discussion and the conclusions will then be generalized.

The state of $C$ at a time $n_{1}\geq n_{i}$ before the specimen may attain one of the alternatives in the first complete set $\{A_{1},A_{2}\}$ may be expressed as follows:

\begin{equation}
\overline{PK_{C}}(n_{1})=\left[\left(c_{1}|A_{1}\right)\ \ \left(c_{2}|A_{2}\right)\right]\:\left[\left(c_{1}c_{11}\ \;c_{2}c_{21}|A_{1}'\right)\ \ \left(c_{1}c_{12}\ \;c_{2}c_{22}|A_{2}'\right)\right].
\label{startstate}
\end{equation}
Here, two juxtaposed quantities $c_{x}$ and $c_{y}$ have a logical or temporal relation in the network of alternatives – they represent propensities in a sequence of possible events. Two groups of such quantities that are separated by a void represent parallel sequences of possible events in the network of alternatives. The groups of this kind that represent propensities that can logically contribute to the propensity of an alternative are associated with that alternative according to Eq. (\ref{association}).

Groups of quantities within the same square bracket represent alternatives in a complete set $\{A_{j}\}$ that cannot be excluded, given the potential knowledge at the given time. Groups of quantities within different square bracket represent alternatives in different complete sets $\{A_{j}\}$ and $\{A_{j'}'\}$.

\subsubsection{Attribute A has knowability level 3}
\label{level3}

At some time $n_{2}>n_{1}$ after one of the alternatives in the complete set $\{A_{1},A_{2}\}$ has been realized, but before the specimen may attain any of the alternatives in the second complete set $\{A_{1}',A_{2}'\}$, the state of the context may be expressed as

\begin{equation}
\overline{PK_{C}}(n_{2})=
\left\{\begin{array}{ll}
\left[A_{1}\right]\:\left[\left(c_{11}|A_{1}'\right)\ \ \left(c_{12}|A_{2}'\right)\right] & or\\
\left[A_{2}\right]\:\left[\left(c_{21}|A_{1}'\right)\ \ \left(c_{22}|A_{2}'\right)\right] &
\end{array}\right.
\label{middlestate3}
\end{equation}
The quantities $c_{1}$ and $c_{2}$ representing the propensities of the alternatives $A_{1}$ and $A_{2}$, respectively, are removed since one of these alternatives is already observed. Given that observation, the probabilities to see the subsequent alternatives $A_{1}'$ or $A_{2}'$ are given by the representations of the corresponding conditional propensities $\pi_{jj'}$. 

At a still later time $n_{3}$, after the final observation of $A'$ at time $n_{f}$, the state will be

\begin{equation}
\overline{PK_{C}}(n_{3})=
\left\{\begin{array}{ll}
\left[A_{1}\right]\:\left[A_{1}'\right] & or\\
\left[A_{1}\right]\:\left[A_{2}'\right] & or\\
\left[A_{2}\right]\:\left[A_{1}'\right] & or\\
\left[A_{2}\right]\:\left[A_{2}'\right] & \\
\end{array}\right.
\label{finalstate3}
\end{equation}

\subsubsection{Attribute A has knowability level 1}
\label{level1}

Just after the time at which the alternatives in the first complete set $\{A_{1},A_{2}\}$ are attained, it is possible to write

\begin{equation}
\overline{PK_{C}}(n_{2})=\left[A_{1}\ \ A_{2}\right]\:\left[\left(c_{1}c_{11}\ \;c_{2}c_{21}|A_{1}'\right)\ \ \left(c_{1}c_{12}\ \;c_{2}c_{22}|A_{2}'\right)\right]
\label{middlestate1}
\end{equation}
for some $n_{2}<n_{f}$, where $n_{f}$ is the time at which one of the alternatives in the second complete set $\{A_{1}',A_{2}'\}$ is observed. That both alternatives $A_{1}$ and $A_{2}$ remain within the first square bracket represents the fact that it is unknowable which of them has occured.

After the final measurement, it holds that

\begin{equation}
\overline{PK_{C}}(n_{3})=
\left\{\begin{array}{ll}
\left[A_{1}\ \ A_{2}\right]\:\left[A_{1}'\right] & or\\
\left[A_{1}\ \ A_{2}\right]\:\left[A_{2}'\right]. &
\end{array}\right.
\label{finalstate1}
\end{equation}

\subsubsection{Delayed observation of attribute A}
\label{delayedchoice}

The event that defines the attainment of attribute $A$ occurs before the attainment of attribute $A'$. Even so, the actual observation of one of the alternatives $\{A_{1},A_{2}\}$ may occur afterward, in a so called \emph{delayed-choice experiment} \cite{delayed,eraser1,eraser2,eraser3}. In that case, the initial state according to Eq. (\ref{startstate}) remains at some time $n_{2}$ after the attainment of $A$, until the observation of attribute $A'$:

\begin{equation}
\overline{PK_{C}}(n_{2})=\left[\left(c_{1}|A_{1}\right)\ \ \left(c_{2}|A_{2}\right)\right]\:\left[\left(c_{1}c_{11}\ \;c_{2}c_{21}|A_{1}'\right)\ \ \left(c_{1}c_{12}\ \;c_{2}c_{22}|A_{2}'\right)\right].
\label{delayedstate}
\end{equation}

Just after the observation of $A'$, the state becomes

\begin{equation}
\overline{PK_{C}}(n_{3})=
\left\{\begin{array}{ll}
\left[\left(c_{1}|A_{1}\right)\ \ \left(c_{2}|A_{2}\right)\right]\:\left[A_{1}'\right] & or\\
\left[\left(c_{1}|A_{1}\right)\ \ \left(c_{2}|A_{2}\right)\right]\:\left[A_{2}'\right] &
\end{array}\right.
\label{finalstate2}
\end{equation}

When the experiment ends by the observation of $A$, the state reduces to that expressed in Eq. (\ref{finalstate3}). If attribute $A$ has knowability level 2 at the start of the experiment, and is still unobserved when the experiment ends even though the possibility to observe it remains, then the state expressed in Eq. (\ref{finalstate2}) may hold indefinitely.

\subsection{Choice of algebraic model of the formalism}
\label{algebraicform}

The above formalism adds little value when $A$ has knowability level 3 and is observed before $A'$. Given the initial state in Eq. (\ref{startstate}), the probability to see alternative $A_{j}$ at some time $n_{1}$ is provided as

\begin{equation}
\left(c_{j}|A_{j}\right)\Rightarrow p_{j}=f(c_{j}).
\label{prob1}
\end{equation}

Given the observation of $A_{j}$, Eq. (\ref{middlestate3}) then gives the probability

\begin{equation}
\left(c_{jj'}|A_{j}'\right)\Rightarrow p_{jj'}=f(c_{jj'})
\label{prob2}
\end{equation}
for the subsequent observation of alternative $A_{j'}'$. The classical rules of probability make it possible to calculate the total probability to observe $A_{j'}'$ as

\begin{equation}
p_{j'}'=f(c_{1})f(c_{1j'})+f(c_{2})f(c_{2j'}),
\end{equation}
in accordance with Eq. (\ref{classicalp}).

To obtain the probability for the observation of $A_{j'}'$ in the delayed-choice case described in to section \ref{delayedchoice}, it is not possible to calculate probabilities sequentially according to Eqs. (\ref{prob1}) and (\ref{prob2}). The relevant state from which this probability should be extracted is rather that expressed in Eq. (\ref{delayedstate}), meaning that

\begin{equation}
\left(c_{1}c_{1j'}\ \ c_{2}c_{2j'}|A_{j'}'\right)\ \Rightarrow \ p_{j'}'=f(c_{1})f(c_{1j'})+f(c_{2})f(c_{2j'}).
\label{delayedp}
\end{equation}

When attribute $A$ has knowability level 1, probabilities should instead be extracted from the state shown in Eq. (\ref{middlestate1}), so that

\begin{equation}
\left(c_{1}c_{1j'}\ \ c_{2}c_{2j'}|A_{j'}'\right)\ \Rightarrow \ p_{j'}'=f(\mathcal{M}\{c_{1}c_{1j'}\ \ c_{2}c_{2j'}\}),
\label{finalp1}
\end{equation}
where $\mathcal{M}\{X\}$ denotes an algebraic model of $X$. Note that it is \emph{not} possible in this case to express $p_{j'}'$ as in Eq. (\ref{delayedp}), since the quantities $f(c_{1})$, $f(c_{1j'})$, $f(c_{2})$ and $f(c_{2j'})$ do not correspond to any probabilities, so that the classical laws used in this equation to combine probabilities do not apply.

It is clear that $\mathcal{M}\{c_{1}c_{1j'}\ \ c_{2}c_{2j'}\}$ must be a quantity of the same algebraic kind as each individual quantity $c_{x}$, since it should be possible to apply the same function $f$ to both of them. Therefore, it is reasonable to require that they belong to the same algebraic \emph{ring}. In that case, the juxtaposition of $c_{j}c_{jj'}$ and the gap between the two pairs of quantities of this kind should be modelled in terms of multiplication ($\times$) and addition ($+$).

Comparing the expressions for $p_{j'}'$ in Eqs. (\ref{delayedp}) and (\ref{finalp1}), it is very natural to choose an algebraic model of the formalism such that

\begin{equation}
\mathcal{M}\{c_{1}c_{1j'}\ \ c_{2}c_{2j'}\}=c_{1}\times c_{1j'}+c_{2}\times c_{2j'}.
\label{model}
\end{equation}

This choice can be further motivated as follows. Consider a hypothetical quantity $c_{x}$ that corresponds to an alternative, or a relation between alternatives, which is not defined or cannot be attained. It can be excluded from the state representations in section \ref{symbolic}, and in the corresponding rules to calculate probabilities. Consequently, if any of the juxtaposed quantities $c_{x}c_{y}c_{z}\ldots$ in a sequence of conceivable events that are causally related correspond to such an alternative, or relation between alternatives, which cannot occur, then the entire sequence can be excluded from the symbolic representation of the state of context $C$. If this were not the case, it would be necessary to include an infinitude of hypothetical chains of alternatives in any symbolic state representation.

A proper algebraic model of these circumstances is

\begin{equation}
\pi_{x}=0\Leftrightarrow c_{x}=0
\label{nozero}
\end{equation}
for an arbitrary propensity $\pi_{x}$, and

\begin{equation}
c_{x}=0\Rightarrow\mathcal{M}\{c_{x}c_{y}c_{z}\ldots\}=0,
\end{equation}
and likewise for $c_{y}$, $c_{z}$ or any other quantity in the sequence. This condition is obviously fulfilled in the proposed algebraic model (\ref{model}), in which

\begin{equation}
\mathcal{M}\{c_{x}c_{y}c_{z}\ldots\}=c_{x}\times c_{y}\times c_{z}\times\ldots.
\end{equation}

On the other hand, the conclusion that a given sequence of conceivable alternatives $c_{x}c_{y}c_{z}\ldots$ can be omitted from the symbolic state representation does not imply that any other such sequence $c_{x}'c_{y}'c_{z}'\ldots$ can be omitted. Therefore, it must hold that

\begin{equation}
\mathcal{M}\{c_{x}c_{y}c_{z}\ldots\}=0\not\Rightarrow\mathcal{M}\{c_{x}c_{y}c_{z}\ldots\;\; c_{x}'c_{y}'c_{z}'\ldots\}=0,
\end{equation}
so that the void between the sequences of juxtaposed quantities cannot be modelled as a multiplication, but is naturally modelled as an addition, like in Eq. (\ref{model}), so that

\begin{equation}
\mathcal{M}\{c_{x}c_{y}c_{z}\ldots\;\; c_{x}'c_{y}'c_{z}'\ldots\}=c_{x}c_{y}c_{z}\ldots+ c_{x}'c_{y}'c_{z}'\ldots.
\end{equation}

Given this algebraic model, it is concluded from Eqs. (\ref{finalp1}) and (\ref{model}) that

\begin{equation}
p_{j'}'=f(c_{1}c_{1j'}+c_{2}c_{2j'})
\label{plevel1}
\end{equation}
in the case where attribute $A$ has knowability level 1. Here, juxtapositions are again understood as multiplications, according to convention.

The philosophical assumption of ontological closure, discussed in section \ref{closure}, implies that proper physical models must distinguish between those contexts $C$ for which there is path information, and those for which there is not. This implies that the recipe to calculate probabilities conditioned on propensities that are probabilites themselves should be different from the recipe where these propensities are not probabilities, as expressed in Eqs. (\ref{classicalp}) and (\ref{quantump}). Therefore,

\begin{equation}
f(c_{1}c_{1j'}+c_{2}c_{2j'})\neq f(c_{1})f(c_{1j'})+f(c_{2})f(c_{2j'}).
\label{quantumdiff}
\end{equation}
This fact excludes the choice $f(x)=x$ for all $x$. That is, $f$ cannot be the identity function:

\begin{equation}
f\neq I.
\label{noid}
\end{equation}

Consider a context $C$ as described in section \ref{level1}, where attribute $A$ has knowability level 1, but suppose that $\pi_{21}=0$, so that observation of alternative $A_{1}'$ implies alternative $A_{1}$ of $A$. According to Eq. (\ref{plevel1}), it holds that $p_{1}'=f(c_{1}c_{11})$. However, the situation is ambiguous in terms of knowability. If $A_{1}'$ is indeed observed, it is immediately known that alternative $A_{1}$ has occurred.

In a modified context $C'$ according to section \ref{delayedchoice} in which $A$ is actually observed, but after $A'$, observation of $A_{1}'$ means that alternative $A_{1}$ will necessarily be observed. However, in this case the probability $p_{1}$ is defined, and hence $p_{11}$, so that it becomes possible to write $p_{1}'=f(c_{1})f(c_{11})$. Since the probability to observe $A_{1}'$ should be the same in contexts $C$ and $C'$, the function $f$ must clearly be such that

\begin{equation}
f(xy)=f(x)f(y)
\label{productf}
\end{equation}
for all choices of $x$ and $y$.

The set of candidate functions $f$ can be further restricted by noting that when attribute $A$ has knowability level 3, it must hold that $p_{1}+p_{2}=1$, as well as $p_{11}+p_{12}=1$ and $p_{21}+p_{22}=1$, so that

\begin{equation}\begin{array}{lll}
1 & = & f(c_{1})+f(c_{2})\\
1 & = & f(c_{11})+f(c_{12})\\
1 & = & f(c_{21})+f(c_{22}).
\end{array}
\label{ccond1}
\end{equation}

When attribute $A$ has knowability level 1, the probabilities $p_{j}$ and $p_{jj'}$ are not defined, but it still holds that $p_{1}'+p_{2}'=1$. Therefore,

\begin{equation}\begin{array}{lll}
1 & = & f(c_{1}c_{11}+c_{2}c_{21})+f(c_{1}c_{12}+c_{2}c_{22})
\end{array}
\label{ccond2}
\end{equation}
according to Eq. (\ref{plevel1}).

It is assumed in section \ref{algebraic} that there is a single algebraic scheme that can describe a wide class of experimental contexts $C$. In particular, it is assumed in section \ref{allknow} that the same rule to calculate probabilities should apply regardless the knowability levels of the involved complete sets of alternatives. Therefore, a function $f$ that expresses such a general rule must fulfill both Eq. (\ref{ccond1}) and Eq. (\ref{ccond2}), even though these equations are derived under mutually exclusive conditions with regard to the knowability level of the set of alternatives associated with attribute $A$.

It should be noted that in a purely classical treatment of probability, corresponding to the choice $f(x)\equiv x$, Eq. (\ref{ccond2}) follows from Eq. (\ref{ccond1}). However, whenever $f(x)\neq x$ for some $x$, as required by Eq. (\ref{noid}), each of the four conditions expressed in these two equations add an independent restriction on the function $f$.

From this fact it can immediately be concluded that no function with real arguments is acceptable. That is, $f:\mathbb{R}\rightarrow\mathbb{R}$ is ruled out. If all quantities $c_{x}$ are always real, Eqs. (\ref{ccond1}) and (\ref{ccond2}) give four conditions that relate six real numbers. This means that there are two independent quantities $c_{j}$ and $c_{jj'}$, and therefore two independent corresponding propensities. However, the number of independent propensities that the function $f$ should be able to generate is $MM'-1=3$, according to section \ref{allprob}.

Let us therefore investigate if there are any acceptable functions $f$ with complex arguments, that is $f:\mathbb{C}\rightarrow\mathbb{R}$.

There are two operations on a pair of complex numbers that have property (\ref{productf}), namely complex conjugation and exponentiation. Therefore it holds that $f(x)=(x^{*})^{\gamma}x^{\eta}$. Since $f(x)$ is real, it must be required that $\gamma=\eta$, so that

\begin{equation}
f(x)=|x|^{2\gamma},
\label{exponentialf}
\end{equation}
for some real $\gamma$. 

To restrict the set of candidate functions $f$ even further, the requirements expressed in sections \ref{allprob} and \ref{independent} can be used. As a check whether these requirements are reasonable, it should be investigated whether they are fulfilled by the choice $\gamma=1$, that is, whether Born's rule $f(x)=|x|^{2}$ passes the test.

The conditions (\ref{ccond1}) and (\ref{ccond2}) become

\begin{equation}\begin{array}{lll}
1 & = & |c_{1}|^{2}+|c_{2}|^{2}\\
1 & = & |c_{11}|^{2}+|c_{12}|^{2}\\
1 & = & |c_{21}|^{2}+|c_{22}|^{2}\\
0 & = & 2\mathrm{Re}\{c_{1}c_{2}^{*}(c_{11}c_{21}^{*}+c_{12}c_{22}^{*})\},
\end{array}
\label{c2cond}
\end{equation}
where $\mathrm{Re}\{x\}$ denotes the real part of $x$.

The first line in this equation corresponds to one condition relating four real quantities $(u_{1},v_{1},u_{2},v_{2})$ defined by $c_{j}=u_{j}+iv_{j}$. There are therefore three independent quantities available to parametrize the first complete set of alternatives $\{A_{1},A_{2}\}$. The minimum number required according to section \ref{allprob} in order to be able to choose propensities freely is $M-1=1$.

The second and third line of Eq. (\ref{c2cond}) correspond to two conditions relating eight real quantities $\{u_{jj'},v_{jj'}\}$ defined by $c_{jj'}=u_{jj'}+iv_{jj'}$. In order to respect the independence of parameters describing different complete sets of alternatives $\{A_{j}\}$ and $\{A_{j'}'\}$ according to section \ref{independent}, it must be possible to choose the complex quantities $\{c_{jj'}\}$ independently from $\{c_{j}\}$. In order to guarantee that the last condition in Eq. (\ref{c2cond}) is always fulfilled, it must therefore hold that

\begin{equation}\begin{array}{lll}
0 & = & c_{11}c_{21}^{*}+c_{12}c_{22}^{*}.
\end{array}
\label{criticalcond}
\end{equation}
This equation corresponds to two more conditions relating the eight real quantities $\{u_{jj'},v_{jj'}\}$. In total, there are thus four such conditions, and therefore four independent quantities available to parametrize the second complete set of alternatives $\{A_{j'}'\}$. The minimum number required according to section \ref{allprob} in order to be able to choose propensities freely is $M(M'-1)=2$.

Therefore, Born's rule is an acceptable general rule to calculate probabilities according to the requirements formulated in this paper, at least for contexts similar to Mach-Zehnder interferometers, for which $M=M'=2$.

What about other choices of $f(x)$? Given Eq. (\ref{exponentialf}), the function can be expressed as a Taylor series according to

\begin{equation}
f(x)=1+\gamma(|x|^{2}-1)+\frac{\gamma(\gamma-1)}{2}(|x|^{2}-1)^{2}+\frac{\gamma(\gamma-1)(\gamma-2)}{6}(|x|^{2}-1)^{3}+\ldots
\label{taylor}
\end{equation}

The next simplest choice after $f(x)=|x|^{2}$ is $f(x)=|x|^{4}$. This function can be used to familiarize oneself with the effect of terms of higher order than two in Eq. (\ref{taylor}), when it comes to the ability of $f(x)$ to fulfill the requirements. In this case, instead of Eq. (\ref{criticalcond}), it follows from Eq. (\ref{ccond2}), and the condition that it should be possible to choose $\{c_{jj'}\}$ independently from $\{c_{j}\}$, that the following conditions must be fulfilled.

\begin{equation}\begin{array}{lll}
0 & = & c_{11}c_{21}c_{21}^{*}c_{21}^{*}+c_{12}c_{22}c_{22}^{*}c_{22}^{*}\\
0 & = & c_{11}c_{11}c_{21}^{*}c_{21}^{*}+c_{12}c_{12}c_{22}^{*}c_{22}^{*}\\
0 & = & c_{11}c_{11}^{*}c_{21}c_{21}^{*}+c_{12}c_{12}^{*}c_{22}c_{22}^{*}\\
0 & = & c_{11}c_{11}c_{11}^{*}c_{21}^{*}+c_{12}c_{12}c_{12}^{*}c_{22}^{*}.
\end{array}\label{messycond}
\end{equation}

These equations express eight conditions relating eight real quantities $\{u_{jj'},v_{jj'}\}$. Adding the two conditions given by the second and third line of Eq. (\ref{c2cond}), it becomes clear that it is impossible to represent a context like that in Fig. \ref{Figure3}(a) by means of complex quantities $\{c_{j},c_{jj'}\}$ when the rule $p_{x}=|c_{x}|^{4}$ is used to calculate corresponding probabilities.

The more higher order terms are included in Eq. (\ref{taylor}), the more conditions like those in Eq. (\ref{messycond}) are introduced. It is concluded that $f(c_{x})=|c_{x}|^{2}$ is the only acceptable choice of function $f$ to calculate probabilities $p_{x}=f(c_{x})$ in experimental contexts $C$ of the Mach-Zehnder type shown in Fig. \ref{Figure3}(a).

Since all other functions than $f(x)=|x|^{2}$ are excluded for this type of context, it is the only one that has the \emph{potential} to be generally applicable in all kinds of experimental contexts. However, it may be asked whether there are other types of contexts for which Born's rule is not applicable, given the conditions expressed in section \ref{assu}.

To answer the question whether there is a sufficient number of independent real parameters in the formalism of section \ref{symbolic} in order to describe a general experimental context $C$, it is sufficient to consider the complete sets of alternatives $\{A_{j'}'\}$ within $C$ one by one, asking whether there is a sufficient number of such parameters to describe each alternative $A_{j'}'$ in this set freely. The reason is the assumed independence of the parts of $C$ that define different sets of alternatives, as specified in section \ref{independent}.

The relevant real parameters are provided by the set of complex quantities $c_{jj'}$, representing the propensity for alternative $A_{j'}'$ given an alternative $A_{j}$ in the complete set $\{A_{j}\}$ that precedes $\{A_{j'}'\}$ in $C$. It does not add anything new to the discussion to introduce more complete sets of alternatives than two, like in the set of contexts represented by Fig. \ref{Figure3}(c).

Suppose that there are $M$ alternatives $A_{j}$ in the complete set $\{A_{j}\}$ and $M'$ alternatives in the complete set $\{A_{j'}'\}$. Clearly, $(M,M')=(2,3)$ for the contexts shown in Fig. \ref{Figure3}(b), and $(M,M')=(3,2)$ for the contexts in Fig. \ref{Figure3}(d).

There are $2MM'$ real quantities $\{u_{jj'},v_{jj'}\}$ defined by $c_{jj'}=u_{jj'}+iv_{jj'}$. Given the choice $f(c_{x})=|c_{x}|^{2}$ for complex $c_{x}$, Eq. (\ref{ccond2}) gives $M(M-1)$ conditions relating these real quantities, generalizing Eq. (\ref{criticalcond}) that holds for $M=M'=2$. The condition $\sum_{j'=1}^{M'}|c_{jj'}|^{2}=1$ for each $j$ provides $M$ new conditions, generalizing the second and third line of Eq. (\ref{c2cond}) in the case $M=M'=2$.

The number of independent real quantities $\{u_{jj'},v_{jj'}\}$ is therefore $2MM'-M(M-1)-M$. According to section \ref{allprob}, the necessary number of independent real quantities in order to represent independent conditional propensities $\pi_{jj'}$ is $M(M'-1)$. Born's rule acting on complex arguments is therefore adequate if and only if

\begin{equation}
M'\geq M-1.
\label{borncondition}
\end{equation} 

This condition means that it is not possible to apply Born's rule to contexts in which the number of possible alternatives decreases too much from one complete set of alternatives to the next. One example would be the type of contexts depicted in Fig. \ref{Figure3}(d) when a fourth alternative $A_{4}$ is added.

The constraint is less severe than it may seem, however. It is often possible to add hypothetical alternatives $\tilde{A}_{\tilde{j}'}'$ to the set $\{A_{j'}'\}$ so that the total number of alternatives $\tilde{M}'$ in the expanded set $\{A_{j'}',\tilde{A}_{\tilde{j}'}'\}$ matches the number $M$ of alternatives $A_{j}$. Consider, for example, a multiple-slit experiment with $M$ slits, and a detector screen with such a poor resolution that $M'<M-1$. There are indeed many more positions $\tilde{A}_{\tilde{j}'}'$ on the screen at which the specimen may end up than there are alternatives defined by the detector. The potential knowledge $PK_{C}(n)$ of the context is often such that it is possible to calculate propensities $\tilde{\pi}_{j\tilde{j}'}$ for these final positions of the specimen, even though they do not correspond to probabilities, since they are not associated with any potential observation.

These considerations show that the set of quantities $\{c_{jj'}\}$ must in some circumstances be seen as a field, being defined for alternatives $\tilde{A}_{\tilde{j}'}'$ that are not explicitly defined or measured within the experimental context. This description matches the conventional picture of a wave function $\Psi(a')$ that permeates the domain of all those attribute values $a'$ that are possible in theory, such as all spatial positions.

\section{Summary and conclusions}
\label{conclusions}

\subsection{Overview}
This paper offers a brief review of previous attempts to derive Born's rule. It is pointed out that it is logically impossible to remove postulates like Born's rule from quantum mechanics without replacing them with other postulates. However, the new postulates may have a different character than the old ones. A trend is identified where one or several formal postulates, like Born's rule, are replaced with operational postulates that describe the process of acquiring information by means of experiment, or by epistemic postulates that describe what the experimenter may know about the experiment. The basic idea of this paper is to follow this trend to its natural end, motivating Born's rule by epistemic assumptions alone, without making any reference to any formal postulates of quantum mechanics.

According to the crucial epistemic assumption, the structure of knowledge and the structure of the physical world reflect each other. This presumed circumstance may be called \emph{ontological closure}. On the one hand, it means that a proper model of the physical world does not contain any elements or make any distinctions that are subjectively unknowable. On the other hand, it means that a proper physical model contains all fundamental forms of appearances and knowable distinctions.

It is claimed that our potential knowledge is always incomplete. In a double-slit experiment, this means that knowledge about which slit the particle passes may forever be unknowable. In that case, a proper physical model should not have hypothetical precise paths as elements, according to the principle of ontological closure. As a consequence, probabilities should not be calculated \emph{as if} there were such paths. Also, the model should distinguish between the two situations in which there is path knowledge, and in which there is not. This means that probabilities should be calculated differently in the two cases.   

In accordance with the above philosophy, \emph{probabilities} are only defined in the present treatment in well-defined experimental contexts, where the outcome to which the probability is associated can actually be observed. Nevertheless, it is often possible to assign \emph{propensities} to outcomes more broadly, by means of symmetry arguments or by application of physical law, regardless whether this outcome can actually be observed. In the double-slit experiment, for example, it is clear that the propensities that the particle passes either slit are equal if the experimental arrangement is symmetrical, regardless whether it can ever be known which slit the particle actually passes. The crucial situation in which it is argued that Born's rule is needed is when there are probabilities that are conditioned on propensities that are not probabilities themselves. 

The above considerations imply that the classical rules of probability cannot be applies in such cases, since they presuppose conditional probabilities referring to precise paths. To pinpoint Born's rule among all conceivable alternative rules, it is assumed that there is a generally applicable physical law. In the present setting this means that there should be an algebraic scheme that can be used to calculate probabilities in all kinds of experimental contexts. The alternatives to Born's rule provide too little parametric freedom to this end.

\subsection{The main argument}
The crucial technical steps in the present motivation of Born's rule are the following. The epistemic state of the experimental context $C$ expressed in Eq. (\ref{middlestate3}) after an attribute $A$ has been attained, but before the subsequent attribute $A'$ is attained, is different from that in Eq. (\ref{middlestate1}). In the former case one alternative $A_{j}$ that increases the knowledge of the value of $A$ is observed, whereas in the latter case it remains unknowable in principle which alternative was attained, and no knowledge about its value is gained. The assumption of ontological closure, discussed in section \ref{closure}, implies that the probabilities for the alternatives $\{A_{j'}'\}$ associated with $A'$ must be calculated differently in the two cases, as expressed by Eqs. (\ref{classicalp}) and (\ref{quantump}).

The assumption that the same general rule to calculate probabilities should nevertheless be applicable in both cases leads to a new normalization condition (\ref{ccond2}), in addition to those in Eq. (\ref{ccond1}), on the function $f(x)$ that expresses this rule. The function $f(x)$ is assumed to be such that its argument $x$ is a representation of any propensity $\pi_{x}$ that is also a probability $p_{x}$.

The choice $f(x)=x$ is excluded since then the new normalization condition follows from the three classical ones, so that the two cases are treated in the same way, contrary to assumption. The extra normalization condition (\ref{ccond2}) also excludes real representations $x$ of propensities, since the number of free parameters becomes too small to express the necessary degree of experimental freedom specified in section \ref{allprob}.

The introduction of complex such representations $c_{x}$ doubles the number of free real parameters, making Born's rule $f(c_{x})=|c_{x}|^{2}$ an acceptable choice. However, given the assumption expressed in section \ref{independent}, any deviation from Born's rule means that new conditions that relate the quantities $\{c_{jj'}\}$ are introduced, like those in Eq. (\ref{messycond}), again shrinking the set of independent parameters too much.

It is not proved that an algebraic scheme involving Born's rule can be applied in all relevant experimental contexts $C$. However, no counterexample has been found. Other acceptable algebraic schemes might be found if some assumptions are relaxed, such as Eqs. (\ref{pf}) and (\ref{nozero}), or the assumption that propensities are represented by real or complex numbers. The existence of such an alternative scheme, which gives rise to different empirical predictions than that involving Born's rule, would show that the present epistemic assumptions, together with the assumption that there is a generally applicable algebraic scheme, are insufficent to pinpoint correct physical law.

\section{Outlook}
\label{outlook}

The epistemic approach to physics employed in this paper is a means to an end, namely to motivate Born's rule. It is not promoted in and of itself. However, since it is argued that this philosophical ansatz is fruitful for this specific purpose, it may be worthwhile to try to use it to make sense of modern physics more broadly.
 
Along this line, consider the symbolic representation $\overline{PK_{C}}$ of the epistemic state of an experimental context $C$, which was introduced in section \ref{symbolic} in order to clarify the specific qualities of such contexts that are essential in the motivation of Born's rule. It is possible to translate key parts of this formalism to states and operations in complex Hilbert spaces, akin to those used in standard quantum mechanics.

To a complete set of $M$ alternatives $\{A_{j}\}$ it is possible to associate such a Hilbert space $\mathcal{H}_{A}$ with dimension $M$, where each alternative $A_{j}$ in this set corresponds to a unit vector $|j\rangle$. The set $\{|j\rangle\}$ can be treated as an orthonormal basis for $\mathcal{H}_{A}$, so that $\langle j|k\rangle=\delta_{jk}$ for the inner product $\langle\cdot|\cdot\rangle$ defined in $\mathcal{H}_{A}$. The association (\ref{association}) between the complex quantity $c_{j}$ and $A_{j}$ corresponds to the element $c_{j}|j\rangle$ in $\mathcal{H}_{A}$. The epistemic state $\left[\left(c_{1}|A_{1}\right)\ \ \left(c_{2}|A_{2}\right)\right]$ in Eq. (\ref{startstate}) translates to a superposition $|A\rangle=c_{1}|1\rangle+c_{2}|2\rangle$. Born's rule gives $\langle A|A\rangle=1$, so that this state vector gets the appropriate unit length. The transition $\left[\left(c_{1}|A_{1}\right)\ \ \left(c_{2}|A_{2}\right)\right]\rightarrow\left[A_{j}\right]$ from the epistemic state of attribute $A$ in Eq. (\ref{startstate}) to that in Eq. (\ref{middlestate3}) when one alternative $A_{j}$ is realized translates to a projection $|A\rangle\rightarrow\bar{P}_{j}|A\rangle=|j\rangle$ in $\mathcal{H}_{A}$. This is an expression of the measurement postulate of standard quantum mechanics.

If two complete sets of alternatives $\{A_{j}\}$ and $\{A_{j'}'\}$ are defined in the same context $C$, it is natural to assign a Hilbert space $\mathcal{H}_{C}=\mathcal{H}_{A}\otimes\mathcal{H}_{A'}$ with dimension $MM'$ to the context as a whole. This is so at least in the case where the corresponding attributes $A$ and $A'$ are simultaneously knowable, that is, when their values are not related according to Heisenberg's uncertainty principle. The final epistemic state $[A_{j}]\:[A_{j'}']$ in Eq. (\ref{finalstate3}) then corresponds to the tensor product $|j\rangle\otimes|j'\rangle$. The set of these product states form an orthonormal basis for $\mathcal{H}_{C}$.

It is a long way to go from here to a construction of quantum mechanics as a whole, but the present perspective on Born's rule unlocks a door for an attempt to achieve that goal. If this is indeed possible, the role of Hilbert spaces and wave functions would be to describe well-defined experimental contexts, rather than describing the state of the world at large.

\end{document}